\documentclass[aps,pre,twocolumn,groupedaddress,showpacs,superscriptaddress]{revtex4}

\usepackage{amsmath}
\usepackage{amssymb}
\usepackage{graphicx}

\begin{document}

\title{Derivation of the phase field crystal model for colloidal solidification}

\author{Sven van Teeffelen}
\email[]{teeffelen@thphy.uni-duesseldorf.de}
 \affiliation{Institut f\"ur Theoretische Physik II, Weiche Materie,
Heinrich-Heine-Universit\"at D\"usseldorf, 
D-40225 D\"usseldorf, Germany}
\author{Rainer Backofen}
\email[]{rainer.backofen@tu-dresden.de}
 \affiliation{Institute of Scientific Computing, 
Technical University Dresden, D-01062 Dresden, Germany}
\author{Axel Voigt}
 \affiliation{Institute of Scientific Computing, 
Technical University Dresden, D-01062 Dresden, Germany}
\author{Hartmut L\"owen}
 \affiliation{Institut f\"ur Theoretische Physik II, Weiche Materie,
Heinrich-Heine-Universit\"at D\"usseldorf, 
D-40225 D\"usseldorf, Germany}
\date{\today}

\begin{abstract} The phase field crystal model is by now widely used
  in order to predict crystal nucleation and growth. For colloidal
  solidification with completely overdamped individual particle
  motion, we show that the phase field crystal dynamics can be derived
  from the microscopic Smoluchowski equation via dynamical density
  functional theory. The different underlying approximations are
  discussed. In particular, a variant of the phase field crystal model
  is proposed which involves less approximations than the standard
  phase field crystal model. We finally test the validity of these
  phase field crystal models against dynamical density functional
  theory.  In particular, the velocities of a linear crystal front
  from the undercooled melt are compared as a function of the
  undercooling for a two-dimensional colloidal suspension of parallel
  dipoles. Good agreement is only obtained by a drastic scaling of the
  free energies in the phase field crystal model in order to match the
  bulk freezing transition point.
\end{abstract}
%
\maketitle
%
%
\section{Introduction}
\label{sec:introduction}
Crystal growth processes are relevant for a variety of different
problems ranging from crystallization of proteins~\cite{Wiener:04} and
other biological macromolecules~\cite{Snell:05} over the construction
of photonic crystals with an optical bandgap~\cite{Dziomkina:05} to
applications in, e.g., metallurgy~\cite{Emmerich:07}.  A full
microscopic understanding of crystal growth with the interparticle
interactions and the thermodynamic boundary conditions as the only
input is still a great challenge since it requires a microscopic
theory of freezing.  Significant progress has been made by using the
so called {\it phase field crystal\/} (PFC) method in which the
traditional phase field theory~\cite{Emmerich:08} is generalized to a
situation with a crystal order parameter.  The PFC model was first
developed by Elder and coworkers~\cite{Elder:02} and then subsequently
applied to many other situations like interfaces~\cite{Athreya:07},
polycrystalline pattern formation~\cite{Wu:07, Goldenfeld:05}, crystal
nucleation~\cite{Backofen:07}, commensurate-incommensurate
transitions~\cite{Achim:06}, and edge dislocations~\cite{Berry:06}.

On the other hand, classical density functional theory (DFT), which
provides a microscopic theory for freezing in
equilibrium~\cite{Oxtoby:91, Singh:91, Loewen:94, Loewen:02, Likos:01,
  Rex1, Rex:06}, was generalized to nonequilibrium situations for
colloidal particles with Brownian dynamics. The so called dynamical
density functional theory (DDFT) can be derived from the basic
Smoluchowski equation~\cite{Archer:04} and was found to be in very
good agreement with Brownian dynamics computer simulations for
dynamics in inhomogeneous situations~\cite{Marconi:99, Dzubiella:03,
  Archer:04} including crystal growth~\cite{SvenPhilMag, SvenPRL}.
The PFC modeling possesses a {\it static\/} part invoking a free
energy expression for order parameters that describe the equilibrium
bulk crystallization transition and a {\it dynamical\/} part that
describes the evolution of the order parameters by a generalized
diffusion or continuity equation.

Recently, Elder and coworkers have derived the static free energy
input to the PFC model from microscopic equilibrium density functional
theory using a truncated density expansion~\cite{Elder:07}. This was
supplemented by a gradient expansion in the order
parameter~\cite{Lowencrystalmult,Lutsko:06}. However, a microscopic
justification and derivation of the dynamical part is still missing.
In this paper, we close this gap for colloidal suspensions where the
individual dynamics is overdamped Brownian motion. Here, the
appropriate microscopic starting point is the Smoluchowski
equation~\cite{Smoluchowski:15,Risken:89} from which by an adiabatic
approximation the dynamical density functional theory can be
derived~\cite{Archer:04}.  In this paper we show that the dynamical
density functional theory provides a firm theoretical basis to derive
the phase field crystal model and to discuss the approximations
involved.  We end up with two versions of the phase field crystal
model, referred to as the PFC1 and PFC2 models, which can both be
implemented numerically with the same effort. We argue that the PFC1
model involves less approximations than the standard PFC2 model, which
is traditionally used in phase field crystal calculations.

Finally we compare the dynamics of freezing for dynamical density
functional theory (DDFT) and phase field crystal (PFC) modeling.  The
system considered here are two-dimensional dipoles with parallel
dipole moments pointing out of their confining plane.  This system is
realized for superparamagnetic colloidal particles which are confined
to a two-dimensional air-water interface at a pending water droplet
and exposed to an external magnetic field~\cite{Zahn:97,
  Haghgooie:05}.  This system is characterized by the pairwise
interaction potential $u(r)=u_0/r^3$, where $u_0$ is a parameter with
dimensions of energy $\times$ volume.  For the specific realization of
two-dimensional paramagnetic colloids of susceptibility $\chi$ exposed
to a perpendicular magnetic field ${\bf B}$, we have $u_0 = (\chi {\bf
  B})^2/2$ in Gaussian units. As for all power-law interactions, the
thermodynamics and structure depend only on one dimensionless coupling
parameter $\Gamma = u_0 \rho^{3/2}/k_BT$, where $\rho$ is the average
one-particle density and $k_BT$ is the thermal energy.

For our comparison between the DDFT and the two PFC models we consider
crystal growth out of an undercooled melt into a two-dimensional
triangular crystal.  The growth velocities of a linear crystal front,
which is cut out of a perfect hexagonal lattice, are calculated as a
function of the undercooling for both the PFC model and the DDFT.
After renormalization of the excess part of the free energy good
agreement is obtained between both approaches implying that the PFC
approach---with appropriately chosen input parameters---provides a
reasonable and justified description framework of crystal growth
phenomena.

The outline of the paper is as follows: In Section~\ref{sec:ddft}, we
derive the DDFT from the Smoluchowski equation of overdamped Brownian
motion, in the same fashion as presented by Archer and
Evans~\cite{Archer:04}. The approximate free energy functional by
Ramakrishnan and Yussouff~\cite{Ramakrishnan:79} is incorporated into
the DDFT in Section~\ref{sec:ry}. Subsequently, in
Section~\ref{sec:pfc}, the two versions of the PFC model (PFC1 and
PFC2) are derived from the approximate form of the DDFT. The different
theories (the DDFT, the PFC1, and PFC2 models) are applied to crystal
growth of dipoles in two dimensions in Section~\ref{sec:application},
including the presentation of the system's equilibrium phase diagram
(Subsection~\ref{subsec:eq}) and of the non-equilibrium problem (the
setup description in Subsection~\ref{subsec:dyn}, and the results in
Subsection~\ref{sec:results}). In Section~\ref{sec:conclusions}, we
summarize and conclude.
%
\section{Dynamical density functional theory (DDFT)}
\label{sec:ddft}
We consider the overdamped dynamics of a set of $N$ identical,
spherical, colloidal particles, immersed in a solvent, which serves
for damping and as a heat bath. Assuming that the particles do
not interact via hydrodynamic forces, the $N$ coupled Langevin
equations of motion~\cite{Risken:89} are given by
\begin{equation}\label{eq:langevin}
\dot{\bf r}_i= \gamma^{-1}\left({\bf F}_i+ {\bf f}_i\right),\quad i=1,\dots,N\,,
\end{equation}
where the dot denotes a time derivative and $\gamma=3\pi\eta_0\sigma$
is the friction coefficient for a colloidal sphere with diameter
$\sigma$ in a fluid of viscosity $\eta_0$. For particles in an
external field $V({\bf r}_i,t)$, which interact with each other via
pairwise additive potentials $u(|{\bf r}_i-{\bf r}_j|)$, the
deterministic force acting on particle $i$ is given by
\begin{equation}\label{systematicforce}
{\bf F}_i(\{{\bf r}\},t)=
-{\bf \nabla}_i\left[\frac{1}{2}\sum_{i,j,i\neq j}u(|{\bf r}_i-{\bf r}_j|)
+ V({\bf r}_i,t)\right]\,,
\end{equation}
where we denote the positions of all particles by $\{{\bf r}\}=\{{\bf
  r}_1,\dots,{\bf r}_N\}$.  The Gaussian white noise random forces
${\bf f}_i$ originating from the solvent are characterized by the
first two moments of their distribution function,
\begin{eqnarray}\label{langevinforcea}
\langle{\bf f}_i(t)\rangle&=& {\bf 0}\\
\label{langevinforceb}
\langle f_{i\alpha}(t) f_{j\beta}(t^\prime)\rangle&=& 
2\gamma k_BT\delta_{ij}\delta_{\alpha\beta}\delta(t-t^\prime)\,,
\end{eqnarray}
where $k_BT$ is the thermal energy. The angle brackets denote a
noise average, and Greek indices indicate a component of the cartesian
vector. Eqs.~\eqref{langevinforcea} and \eqref{langevinforceb} fulfill
the well-known Einstein fluctuation-dissipation relation yielding a
short-time diffusion constant $D=k_BT/\gamma$. The set of coupled,
stochastic differential equations~\eqref{eq:langevin} for the particle
coordinates corresponds to a deterministic Fokker-Planck equation for
the $N$-particle probability density $W(\{{\bf
  r}\},t)$~\cite{Smoluchowski:15,Risken:89},
\begin{eqnarray}\label{eq:smoluchowski}
\dot{W}(\{{\bf r}\},t)=
\mathcal{L}_{\rm S} W(\{{\bf r}\},t)\,,&\\
\label{eq:smoluchowskioperator}
\mathcal{L}_{\rm S}=\gamma^{-1}\sum_i{\bf \nabla}_i\cdot
\left[k_BT{\bf \nabla}_i-{\bf F}_i(\{{\bf r}\},t)\right],&
\end{eqnarray}
which determines the probability to find the set of $N$ particles
within a small volume around the positions $\{{\bf r}\}$ at time $t$,
given a normalized, initial distribution $W(\{{\bf r}\},t=0)$. The sum
runs over all particles $i=1,\dots,N$.
The continuity equation~\eqref{eq:smoluchowski} is
referred to as Smoluchowski equation~\cite{Risken:89}.

For dense, strongly interacting fluids, one is typically not
interested in the position of all individual particles but rather in
the probability to find {\it any} particle at a certain vector ${\bf
  r}$ at time $t$. We therefore introduce the time-dependent one- and
two-particle densities
\begin{eqnarray}\label{eq:rho1}
\rho({\bf r},t)=\sum_{i}\langle\delta({\bf r}-{\bf r}_i(t))\rangle\,,&\\
\label{eq:rho2}\rho^{(2)}({\bf r},{\bf r}',t)=
\sum_{i,j;i\ne j}\langle\delta({\bf r}-{\bf r}_i(t))\delta({\bf r}-{\bf r}_j(t))\rangle\,,&
\end{eqnarray}
where we dropped the superscript ``$(1)$'' on the one-particle
density.  Generally, the $n$-particle density is equal to the
$(N-n)$-times integrated probability density $W$,
\begin{eqnarray}\label{eq:rhon}
\rho^{(n)}({\bf r}_1,\dots,{\bf r}_n,t)
=\frac{N!}{(N-n)!}\int{\rm d}^{N-n}{\bf r}\,W(\{{\bf r}\},t)\,.
\end{eqnarray}
A deterministic equation of motion for the time evolution of
$\rho({\bf r},t)$---Eq.~\eqref{eq:rho1timeevolution}---can on the one
hand be directly derived from the Langevin equations,
Eq.~\eqref{eq:langevin}, via a coordinate transformation ${\bf
  r}_i\rightarrow\hat\rho({\bf r},t)$, where
$\hat\rho({\bf r},t)=\sum_i\delta\left({\bf r}(t)-{\bf r}_i(t)\right)$ is the
one-particle density operator, and a subsequent noise-average.  This
way was followed by Marini Bettolo Marconi and Tarazona
(MT)~\cite{Marconi:99}, based on an earlier approach by
Dean~\cite{Dean:96}. On the other hand, the same equation is obtained
by integrating the Smoluchowski equation~\eqref{eq:smoluchowski} over
the positions of $N-1$ of the $N$ particles and making use of
Eqs.~\eqref{eq:rho1}, \eqref{eq:rho2}, and \eqref{eq:rhon}. The latter
approach was adopted by Archer and Evans~\cite{Archer:04}. The
continuity equation for $\rho({\bf r},t)$ reads
\begin{eqnarray}\label{eq:rho1timeevolution}
\dot{\rho}({\bf r},t)&=&\gamma^{-1}{\bf \nabla}
\cdot\Big[k_BT{\bf \nabla}\rho({\bf r},t) 
+\rho({\bf r},t){\bf \nabla}  V({\bf r},t)\nonumber \\*
&&+\int{\rm d}{\bf r}' \rho^{(2)}({\bf r},{\bf r}',t)\nabla  
u\left(\left|{\bf r}-{\bf r}'\right|\right)\Big]\,.
\end{eqnarray}
For noninteracting particles in zero external field, this equation
reduces to Fick's diffusion equation. Also with an external field
applied, Eq.~\eqref{eq:rho1timeevolution} is exactly solvable. In the
interesting case of interacting particles, however, an expression for
the time-dependent two-particle density $\rho^{(2)}({\bf r},{\bf
  r}',t)$ is still needed.

Within DDFT, $\rho^{(2)}({\bf r},{\bf r}',t)$ is approximated by a yet
unspecified equilibrium two-particle density $\rho^{(2)}_0({\bf
  r},{\bf r}')$; the latter is evaluated at a corresponding
equilibrium fluid, in which the equilibrium density $\rho_0({\bf r})$
is equal to the instantaneous one-particle density $\rho({\bf r},t)$
of the nonequilibrium system.  The approximation of replacing a
time-dependent, nonequilibrium by an equilibrium correlation function,
is referred to as {\it adiabatic} approximation; it goes back to
Enskog~\cite{Enskog:22}, who applied it to the time evolution of the
single-particle distribution function in a dense gas of hard spheres.
In order to render the instantaneous density $\rho({\bf r},t)$ an
equilibrium density, an appropriate external potential $v({\bf r})$
must be applied. That such a potential exists for any physical density
field $\rho({\bf r},t)$, and that it is, further, a unique functional
of the density $\rho_0({\bf r})$, is stated and proved in one of the
basic theorems of density functional theory~\cite{Evans:79}.

The connection to classical density functional theory is now made by
identifying the three different terms in the bracket on the right-hand
side of Eq.~\eqref{eq:rho1timeevolution} with terms of the form
$\rho({\bf r})\nabla\delta F_i[\rho]/\delta\rho({\bf r})$, where the
$F_i[\rho]$ are different contributions to the Helmholtz free energy
functional $F[\rho({\bf r})]$, which is provided by classical density
functional theory.  The functional $F[\rho({\bf r})]$ is a unique
functional of the static one-particle density $\rho({\bf
  r})$~\cite{Evans:79}. If $F[\rho({\bf r})]$ is known exactly it is
minimized by the equilibrium one-particle density $\rho({\bf
  r})=\rho_0({\bf r})$, where it takes the value of the Helmholtz free
energy $F\equiv F[\rho_0({\bf r})]$. The functional is divided into
three terms:
\begin{equation}\label{eq:ftot}
F\left[\rho({\bf r})\right]= F_{\rm{id}}\left[\rho({\bf r})\right]
+F_{\rm{ex}}\left[\rho({\bf r})\right]+F_{\rm{ext}}\left[\rho({\bf
    r})\right]\,.
\end{equation}
The ideal gas part, which is of completely entropic nature and which
yields the (first) diffusion term in Eq.~\eqref{eq:rho1timeevolution},
is
\begin{equation}\label{eq:fid}
F_{\rm id}\left[\rho({\bf r})\right]= k_B T \int\,{\mathrm d} {\bf r} 
\rho({\bf r})\left\{\ln\left[\rho({\bf r})\Lambda^d\right]-1\right\}\,,
\end{equation}
with $\Lambda$ denoting the thermal de Broglie wavelength and $d$ the
spatial dimension. The external part corresponding to the second term
in Eq.~\eqref{eq:rho1timeevolution} is given by
\begin{equation}\label{eq:fext}
  F_{\rm{ext}}\left[\rho({\bf r})\right]=\int\,{\mathrm d} {\bf
    r}\rho({\bf r})V({\bf r},t)\,.
\end{equation}
Finally, the excess part $F_{\rm ex}[\rho({\bf r})]$, originating from
the correlations between the particles, is generally unknown and must
be approximated; a specific approximation is introduced further down.
Note that the excess part is {\it not} the potential energy of
interaction but a contribution to the free energy.  The connection of
the excess part and the third term on the right-hand side of
Eq.~\eqref{eq:rho1timeevolution} is made through the sum rule
\begin{equation}\label{eq:sumrule}
-\rho_0({\bf r}){\bf \nabla} c_0^{(1)}({\bf r})=
(k_BT)^{-1}\int{\rm d}{\bf r}'
\rho^{(2)}_0({\bf r},{\bf r}'){\bf \nabla}u\left(\left|{\bf r}-{\bf r}'\right|\right)\,,
\end{equation}
which connects the two-particle density $\rho^{(2)}_0({\bf r},{\bf
  r}')$ with the effective one-body potential $k_BTc_0^{(1)}({\bf r})$.
The latter, in turn, is---up to a minus sign---equal to the first
functional derivative of the excess free energy functional $F_{\rm
  ex}[\rho_0({\bf r})]$ with respect to density,
\begin{equation}\label{eq:c1}
  k_BTc_0^{(1)}({\bf r})=-\frac{\delta F_{\rm ex}[\rho_0({\bf r})]}{\delta \rho({\bf r})}\,.
\end{equation}
Using Eqs.~\eqref{eq:sumrule} and \eqref{eq:c1}, we can therefore
rewrite Eq.~\eqref{eq:rho1timeevolution} as
\begin{eqnarray}
 \label{eq:ddft1}
 \dot\rho({\bf r},t) &=&
\gamma^{-1}  \bigg\{ k_BT\nabla^2 \rho({\bf r},t)
+\nabla\cdot\left[\rho({\bf r},t)\nabla V({\bf r},t)\right]   \nonumber \\
&& -\nabla \cdot\left[ \rho({\bf r},t)
   \nabla \frac{\delta F_{\rm ex}\left[\rho({\bf r},t)\right]}{\delta \rho({\bf r},t)} 
\right]\bigg\}\,,
\end{eqnarray}
which, making use of Eqs.~\eqref{eq:ftot}, \eqref{eq:fid}, and
\eqref{eq:fext}, reads in a compact form
\begin{equation}\label{eq:ddft}
  \dot\rho({\bf r},t)
  =\gamma^{-1}\nabla \cdot \left[\rho({\bf r},t)\nabla
\frac{\delta F\left[\rho({\bf r},t)\right]}{\delta \rho({\bf r},t)}\right]\,.
\end{equation}
Eq.~\eqref{eq:ddft} constitutes the fundamental, nonlinear,
deterministic equation for the time-evolution of the one-particle
density $\rho({\bf r},t)$ and will be referred to as DDFT equation
henceforth. For time-independent external potentials $V({\bf r})$, the
DDFT describes the relaxation dynamics of the density field towards
equilibrium at the minimum of the Helmholtz free energy functional
$F[\rho_0]$, given an exact canonical excess free energy
functional $F_{\rm ex}[\rho]$. The path in the space of density fields
is in general not the one of steepest descent, but is governed by the
mass conservation constraint in Eq.~\eqref{eq:ddft}~\cite{Marconi:99}.

Eq.~\eqref{eq:ddft} is a deterministic equation; it has no extra
noise-term, as all the fluctuations---given that $F_{\rm ex}[\rho]$ is
exact---are already taken into account. As discussed at large by
MT~\cite{Marconi:99,Marconi:00}, the addition of a noise term in
Eq.~\eqref{eq:ddft1} leads to an overcounting of fluctuations. Archer
and Rauscher have discussed the possibility of including a noise term
if $\rho({\bf r},t)$ is not interpreted as the ensemble averaged but
as a coarse-grained (in time) probability distribution
$\overline{\rho}({\bf r},t)$~\cite{Rauscher}. However, this also
requires a replacement of the functional $F[\rho]$ by a functional of
the coarse-grained probability density, which is {\it not} the
Helmholtz free energy functional of density functional theory. We will
come back to this point in Section~\ref{sec:rauscher}.

The DDFT equation had been suggested earlier on phenomenological
grounds by Evans~\cite{Evans:79} and later by Dieterich {\it et
  al.}~\cite{Dieterich:90}. The same dynamical equation with the
excess free energy functional of Ramakrishnan and
Yussouff~\cite{Ramakrishnan:79,Haymet:81} (cf.\ Section~\ref{sec:ry})
had been derived by Munakata~\cite{Munakata:77,Munakata:77b} and
extended to non-spherical particles in the context of solvation
dynamics by Calef and Wolynes~\cite{Calef:83}, which was later
reformulated by Chandra and Bagchi~\cite{Chandra:89,Chandra:90}; these
equations are referred to as Smoluchowski-Vlasov or {\it nonlinear
  diffusion} equations~\cite{Munakata:77b}, as they are derived from a
Vlasov equation~\cite{resibois-deleener:77} with a Fokker-Planck
collision operator~\cite{Munakata:77b}.  However, MT were the first to
derive the theory from the microscopic equations of motion and to make
clear the contact to static DFT~\cite{Marconi:99,Marconi:00}.  Similar
attempts had been made by other authors before (Kirkpatrick {\it et
  al.}~\cite{Kirkpatrick:89}, Dean~\cite{Dean:96}, and Kawasaki {\it
  et al.}~\cite{Kawasaki:97,Kawasaki:94}), which, however, do not
distinguish the average density $\rho$ with the density operator
$\hat\rho$, which in turn leads to an additional noise term on the
right-hand side of Eq.~\eqref{eq:ddft} and therefore to an
overcounting of fluctuations, given an accurate functional $F[\rho]$
(see also the discussions by MT~\cite{Marconi:99}, Archer and
Rauscher~\cite{Rauscher}, and L\"owen~\cite{Loewen:03}).

The DDFT is an approximate theory in several respects: the first and
most fundamental approximation is the already introduced assumption of
adiabatic relaxation dynamics. In practice, this approximation is most
severe in dynamical processes that are fast compared to the diffusive
time scale of the system. To our knowledge, this issue has been
studied systematically to date only for weak perturbations of a
hard-rod fluid in one dimension by Penna and Tarazona~\cite{Penna:06}.
The use of approximate free energy functionals is the second
fundamental approximation turning out to be severe in many
applications (cf.\ the next section). Third, we did only consider
systems in which hydrodynamic interactions between the particles play
no role.  The latter assumption can be approximately tackled by
allowing for density-dependent friction constants
$\gamma$~\cite{Royall:07}, which is appropriate for long-wavelength
fluctuations of the density field, or by taking hydrodynamic
interactions on the Rotne-Prager (two-particle) level into account, as
was recently demonstrated by Rex and L\"owen~\cite{Rex:08}.
%
\section{Approximate density functionals in the DDFT}
\label{sec:ry}
For most problems including those involving freezing, $F_{\rm
  ex}[\rho({\bf r})]$ is only known
approximately~\cite{Evans:79,Singh:91}; in practice, this restriction
most often constitutes the more severe approximation as compared to
the adiabatic approximation. In this paper, we follow the approach of
Ramakrishnan and Yussouff (RY)~\cite{Ramakrishnan:79} as laid out for
the dipolar system in 2D in reference~\cite{SvenJPCM} and as already
exploited for the DDFT of the same model in
references~\cite{SvenPhilMag, SvenPRL}.  Within the RY-approach,
$F_{\rm{ex}}\left[\rho({\bf r})\right]$ is expanded up to second order
in terms of density difference $\Delta\rho=\rho({\bf r})-\rho$ around
a reference fluid, where the fluid density $\rho$ is chosen the
average density of the inhomogeneous system:
\begin{eqnarray}\label{eq:fex}
F_{\rm{ex}}\left[\rho({\bf r})\right]\simeq F_{\rm{ex}}(\rho)\nonumber \\
-\frac{k_BT}{2}\iint\,{\mathrm d} {\bf r}{\mathrm d}
{\bf r}^\prime \Delta\rho({\bf r})\Delta\rho({\bf
  r}^\prime)c_0^{(2)}({\bf r}-{\bf r}^\prime;\rho)\,.
\end{eqnarray}
Here $F_{\rm{ex}}(\rho)$ and $c_0^{(2)}({\bf r};\rho)$ are the excess
free energy and the direct correlation function of the reference fluid
of density $\rho$, respectively~\cite{pfc_footnote_EMA}.  Despite the
ease of implementation the RY-excess free energy functional is used
here because it will lead directly to the PFC model in the next
section.  Within the RY-approximation the DDFT equation now reads
\begin{align}\begin{split}
 \label{eq:ddft2}
\dot\rho({\bf r},t) =
D  \bigg\{ \nabla^2 \rho({\bf r},t) 
+(k_BT)^{-1}\nabla\cdot\left[\rho({\bf r},t)\nabla V({\bf r},t)\right] \\
 -\nabla \cdot\left[ \rho({\bf r},t)
   \nabla \int {\rm d}{\bf r}' \rho({\bf r}')
 c_0^{(2)}(|{\bf r}-{\bf r}'|; \rho) \right]\bigg\},
\end{split}\end{align}
with $D=k_BT/\gamma$ the diffusion constant.

Apart from leading to quantitatively wrong equilibrium density fields
and free energies the approximate density functional might display
more than one local minimum, in which the system might get trapped.
E.g., in conjunction with freezing a flat and constant density
profile, corresponding to the fluid state, is ``metastable'' at all
temperatures for any known approximate density functional; starting
from the fluid state, within DDFT, the system therefore never reaches
the stable crystalline state which is represented through a
periodically modulated density field. As discussed at length by
MT~\cite{Marconi:99,Marconi:00}, this failure made some groups add a
noise term to Eq.~\eqref{eq:ddft}, which is at best justified {\it a
  posteriori}. In particular, it leads to the already mentioned
overcounting of fluctuations (see also Section~\ref{sec:rauscher}).
%
\section{The Phase Field Crystal (PFC) Model}
\label{sec:pfc}
As the DDFT, the PFC model is based on a free energy functional
$\mathcal F[\psi({\bf r},t)]$ of a phase field $\psi({\bf r},t)$ and a
dynamical equation for the phase field's time evolution similar to the
DDFT equation. The PFC model was introduced as a phenomenological
theory by Elder {\it et al.}~\cite{Elder:02,Elder:04}. If the yet to
be specified functional $\mathcal F[\psi({\bf r},t)]$ is set equal to
a particular approximation of the Helmholtz free energy functional
from density functional theory, i.e., $\mathcal F[\psi({\bf
  r},t)]=F[\rho({\bf r},t)=\psi({\bf r},t)]$, as was also suggested by
Elder {\it et al.}~\cite{Elder:07}, the phase field is consequently to
be interpreted as the density field, i.e., $\psi({\bf r},t)=\rho({\bf
  r},t)$.  We show in this section that the commonly used PFC equation
of motion~\cite{Elder:02,Elder:04} [Eq.~\eqref{eq:PFC2}] can be
regarded as a particularly simplified and further approximated version
of the DDFT with the RY approximation to the excess free energy
functional [cf.\ Eq.~\eqref{eq:ddft2}]; consequently, the PFC model is
derived here from the basic Langevin equations of
motion~\eqref{eq:langevin} for the case of overdamped dynamics.  Of
course, this reasoning only holds if the phase field $\psi({\bf r},t)$
in the PFC model is regarded as the density field $\rho({\bf r},t)$ of
Eq.~\eqref{eq:rho1}, which will be assumed in this section and is
believed to be assumed in many other papers whenever $\mathcal
F[\psi]$ is set equal to $F[\rho]$.  Different interpretations of
$\psi({\bf r},t)$ are discussed in the next section.

Apart from the adiabatic and the RY approximation the derivation goes
via three further approximations; First, the RY excess free energy
functional, Eq.~\eqref{eq:fex}, is approximated by a (local) gradient
expansion. Second, the mobility in the dynamical
equation~\eqref{eq:ddft2}, is set to be constant, i.e.,
$\gamma^{-1}\rho({\bf r},t)\approx\gamma^{-1}\rho$ with $\rho$ the
{\it average} density of the system. Third, the ideal gas part of the
free energy, Eq.~\eqref{eq:fid}, is approximated by its truncated
Taylor series.  According to the (additional) approximations of the
PFC model two different equations of motion are put forward referred
to as the PFC1 model, which is obtained after the first approximation,
and as the PFC2 model, which is obtained after the second and third
approximations.  Obviously, the PFC2 model constitutes an approximate
form of the PFC1 model.

Following the procedure suggested by Elder and
coworkers~\cite{Elder:07}, the RY excess free energy functional,
Eq.~\eqref{eq:fex}, is approximated by its gradient expansion within
both PFC approaches:
\begin{eqnarray} \label{eq:PFC-Fexc}
\mathcal F_{\rm ex}[\rho({\bf r})]&=&F_{\rm ex}(\rho)
- \frac{k_{\rm B}T}{2} 
\int {\rm d {\bf r}} \Delta \rho({\bf r}) 
\Big( \hat{C}_0 - \hat{C}_2 \nabla^2 \nonumber \\*
&&+ \hat{C}_4 \nabla^4
+\dots \Big) \Delta \rho({\bf r}) \,,
\end{eqnarray}
Consequently, $\mathcal F_{\rm ex}$ is {\it local} in the density
field, which renders the yet to be introduced PFC equations of motion
[approximate forms of the DDFT equation~\eqref{eq:ddft2}]
computationally faster to solve.  The gradient expansion is equivalent
to a Taylor-expansion of the Fourier transform $\hat{c}_0^{(2)}({\bf
  k};\rho)$ of the two-particle direct correlation function introduced
in Eq.~\eqref{eq:fex},
\begin{equation} \label{eq:PFC-C2}
 \hat{c}_0^{(2)}({\bf k};\rho)= \hat{C}_0 + \hat{C}_2 k^2+\hat{C}_4k^4 +\dots \,.
\end{equation}
Due to rotational symmetry of the pair correlation function the
expansion is only in even powers of $k$. Truncating the expansion at
fourth order, the time evolution given by Eq.~\eqref{eq:ddft} now reads
\begin{align}\begin{split}\label{eq:PFC1}
\dot\rho({\bf r},t)
  =D\nabla^2 \rho({\bf r},t) +
D\nabla \cdot  \Bigg\{ \rho({\bf r},t) 
\nabla \Big[ (k_BT)^{-1}V({\bf r},t) \\
-\left( \hat{C}_0-\hat{C}_2 \nabla^2 + \hat{C}_4 \nabla^4 \right) 
\rho({\bf r},t) \Big]\Bigg\}\,.
\end{split}\end{align}
This equation, which we refer to as PFC1 model, approximates the
integro-differential equation of the DDFT, Eq.~\eqref{eq:ddft}, by a
local partial differential equation of sixth order. Further down, we
will advocate the use of this equation rather than of the more
approximate equation of the PFC2 model.

The (second) constant mobility approximation, an {\it ad hoc}
assumption of a constant density $\rho({\bf r},t)=\rho$ in front of
the functional derivative in Eq.~\eqref{eq:ddft}, leads to the
equation
\begin{equation}\label{eq:ddftconstmob}
  \dot\rho({\bf r},t)
  =\gamma^{-1}\rho\nabla^2\left[
\frac{\delta {\mathcal F}\left[\rho({\bf r},t)\right]}{\delta \rho({\bf r},t)}\right]\,,
\end{equation}
where the total free energy functional is given by
\begin{equation}
  \label{eq:PFC_Ftot}
\mathcal F[\rho]=F_{\rm id}[\rho]+F_{\rm ext}[\rho]+\mathcal
F_{\rm ex}[\rho]\,.  
\end{equation}
Concurrently, insertion of the ideal gas part of the
Helmholtz free energy functional, Eq.~\eqref{eq:fid}, leads to a term,
which is logarithmic in the density field,
\begin{equation}\label{eq:constmobfidderivative}
\nabla^2\left[\frac{\delta F_{\rm id}\left[\rho({\bf r},t)\right]}{\delta \rho({\bf r},t)}\right]
=k_BT\nabla^2\ln\left[\rho({\bf r},t)\Lambda^d\right]\,.
\end{equation}
This term replaces the simpler diffusion term, $\nabla^2\rho({\bf
  r},t)$, in the DDFT equation (and in the PFC1 model). Within the
PFC2 model the logarithm is expanded in a power series about the
constant density $\rho$, i.e.,
\begin{align}\begin{split}\label{eq:fid_cpfc}
F_{\rm id}\left[\rho({\bf r})\right]\approx k_B T\rho \int\,{\mathrm d} {\bf r} 
\Big\{\frac{1}{2}\phi({\bf r},t)^2-\frac{1}{6}\phi({\bf r},t)^3\\
+\frac{1}{12}\phi({\bf r},t)^4-{\rm const.}\Big\}
\end{split}\end{align}
with $\phi({\bf r},t)=[\rho({\bf r},t)-\rho]/\rho$ the dimensionless
density modulation. This leads to the standard form of the PFC
model used in the literature,
\begin{align}\begin{split}\label{eq:PFC2}
    \dot\phi({\bf r},t) =D \rho \nabla^2 \Bigg[ \phi({\bf r},t)
    - \frac{1}{2}\phi({\bf
      r},t)^2 +  \frac{1}{3} \phi({\bf r},t)^3 \\
    +(k_BT)^{-1}V({\bf r},t) - \rho \left(\hat{C}_0-\hat{C}_2 \nabla^2
      + \hat{C}_4 \nabla^4 \right) \phi({\bf r},t) \Bigg] \,,
\end{split}\end{align}
which is henceforth referred to as the constitutive equation of the
PFC2 model. Note that the second and third term on the right-hand side
only appear due to the constant-mobility assumption and are not
present in the less approximate Eq.~\eqref{eq:PFC1}.

We will use the (standard) PFC2 model, Eq.~(\ref{eq:PFC2}), as well as
the more accurate PFC1 model, Eq.~(\ref{eq:PFC1}), and compare them
with the DDFT, employing the RY approximation, Eq.~\eqref{eq:ddft2}.
Therefore we need to parametrize $\hat{C}_0,\hat{C}_2$, and
$\hat{C}_4$ in the PFC1 and PFC2 models according to $\hat{c}({\bf k};
\Gamma)$ in Eq.~\eqref{eq:PFC-C2}. A particular parametrization,
motivated from the {\it one-mode} approximation to the PFC2
model~\cite{Elder:04}, is chosen and presented in
Section~\ref{sec:application}. Before we come to the comparison we
comment shortly on the use of an additional noise term in
Eqs.~\eqref{eq:PFC1} and \eqref{eq:PFC2} in the following section.
%
\section{Noise term in the PFC equation}
\label{sec:rauscher}
Typically, the PFC2 equation~\eqref{eq:PFC2} is supplemented by a
non-multiplicative Gaussian noise term $\eta({\bf
  r},t)$~\cite{Elder:02} fulfilling
\begin{eqnarray}\label{pfcnoisea}
\langle\eta({\bf r},t)\rangle&=& 0\,,\\*
\label{pfcnoiseb}
\langle\eta({\bf r},t)\eta({\bf r}',t')\rangle &=& 
2 k_BT\gamma^{-1}\delta({\bf r}-{\bf r}')\delta(t-t^\prime)\,.
\end{eqnarray}
As already pointed out in Section~\ref{sec:ddft}, such a noise term
can not be derived in the context of DDFT, since the density field
$\rho({\bf r},t)$ is an ensemble averaged quantity. Instead, the
addition of a noise term in Eq.~\eqref{eq:ddft1} leads to an
overcounting of fluctuations~\cite{Marconi:99}. Therefore we point out
that---at least for colloidal dynamics---the addition of a noise
term is not well justified.

However, Archer and Rauscher~\cite{Rauscher} argue on a
phenomenological basis that a noise term fulfilling
Eqs.~\eqref{pfcnoisea} and \eqref{pfcnoiseb} can be introduced if the
phase field is {\it not} understood as the ensemble averaged
probability density $\rho({\bf r},t)$ but as a coarse-grained (in
time) density
\begin{equation}
\overline{\rho}({\bf r},t)=\int_{-\infty}^t{\mathrm d}t'
K(t-t')\hat\rho({\bf r},t')\,,
\end{equation}
with $K(t)$ a coarse-graining function of width
$\tau=\int_0^\infty\mathrm d t \,t K(t)$ and $\hat \rho({\bf
  r},t)=\sum_i\delta\left({\bf r}(t)-{\bf r}_i(t)\right)$ the density
operator, as already introduced above. However, if this approach is
followed, the functional $F[\rho]$ needs to be replaced by a
functional of the coarse-grained probability density
$\overline{\rho}({\bf r},t)$, which is {\it not} the Helmholtz free
energy functional of density functional theory and which is generally
unknown. Second, the temperature entering Eq.~\eqref{pfcnoiseb} must
be renormalized by a factor $\sqrt{\tau_0/\tau}$ accounting for the
ratio of the microscopic time-scale $\tau_0$ and the coarse-graining
time scale $\tau$. In fact, the dependence of Eq.~\eqref{pfcnoiseb} on
$\tau_0$ points to a conflict with a fundamental assumptions of
Brownian dynamics, namely that $\tau_0$ is much smaller than any
relevant time scale.  If this reasoning is followed nevertheless, the
non-multiplicative nature of the noise-term, Eq.~\eqref{pfcnoisea},
comes about only after an approximation similar to the one of constant
mobility in the PFC2 model whereas the less approximate PFC1 model
should be appended by a multiplicative noise term of the form
\begin{equation}
\eta({\bf r},t)\rightarrow{\bf \nabla}\sqrt{\frac{\tau_0}{\tau}
\overline{\rho}({\bf r},t)}\,\eta({\bf r},t)\,,
\end{equation}
where the gradient assures mass conservation. Due to the gradient the
\^Ito and Stratonovich calculus~\cite{Risken:89} are equivalent.
For a phenomenological motivation of the according dynamical equation,
we refer the reader to Archer and Rauscher~\cite{Rauscher}.  In the
following, we do not consider a fluctuating density field but restrict
our study to the application of the deterministic DDFT, PFC1, and PFC2
equations, respectively. Finally, we remark that a noise term in the
case of molecular dynamics was recently discussed by Tupper and
Grant~\cite{Tupper:08}.
%
%
\section{Application of the DDFT and the PFC model to propagating
  crystal fronts}
\label{sec:application}
In this and the following section, we compare the three different
approaches to the non-equilibrium, Brownian dynamics discussed, the
DDFT and the two PFC models, to the problem of propagating crystal
fronts. For a better understanding of what drives the crystal growth,
we present, first, the equilibrium phase diagram obtained from the
underlying free energy functionals in Subsection~\ref{subsec:eq}
before studying the dynamics in Subsection~\ref{subsec:dyn}.
\subsection{The equilibrium state}
\label{subsec:eq}
\begin{figure}
\begin{center}
  \includegraphics[width=8.5cm]{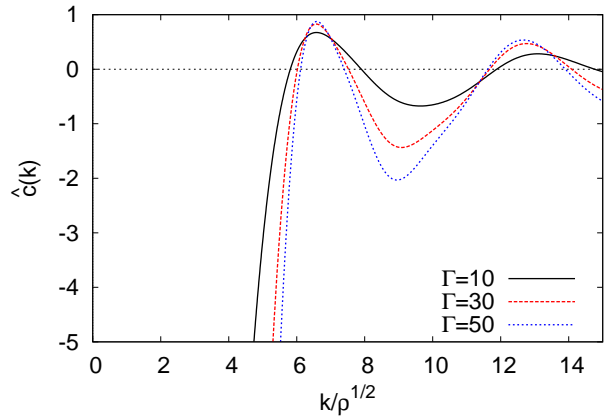}
  \caption{(Color online) The Fourier transform $\hat c(k)$ of the two-particle
    direct correlation function for different coupling constants
    $\Gamma = 10, 30, 50$, plotted against $k/ \rho^{1/2}$. 
  }
  \label{fig:cofk}
\end{center}\end{figure}

Input to the static free energy functionals, Eq.~\eqref{eq:ftot} and
Eq.~\eqref{eq:PFC_Ftot}, and, concurrently, to the dynamical theories is the
direct correlation function of the fluid $c_0^{(2)}({\bf
  r})$~\cite{hansen-mcdonald:86}, which has been obtained for a large range of
coupling constants $0<\Gamma\leq62.5$ from liquid-state integral equation
theory as described in references~\cite{SvenJPCM, Teeffelen:06}. In
particular, $c_0^{(2)}({\bf r})$ has been obtained by iteratively solving the
coupled Ornstein-Zernicke equation~\cite{hansen-mcdonald:86} and the closure
relation suggested by Rogers and Young~\cite{Rogers:84}. The dimensionless
Fourier transform of the pair correlation function $\hat c(k)=\rho\tilde
c_0^{(2)}(k)$ as a function of wave vector is plotted for different coupling
constants $\Gamma$ in Fig.~\ref{fig:cofk}.

Whereas the excess free energy in the RY approximation to the DFT,
Eq.~\eqref{eq:fex}, requires, in general, the complete correlation
function $\hat c(k)$, the respective function in the PFC models,
Eq.~\eqref{eq:PFC-Fexc}, only needs the parameters
$\hat{C}_0,\hat{C}_2$, and $\hat{C}_4$ as an input. The latter can in
principle be obtained from $\hat c(k)$ in different ways. For our
purpose they are chosen according to a series expansion of $\hat c(k)$
in terms of $k^2$ about the correlation function's first maximum at
$k=k^*$ up to second order in $k^2$, i.e.,
\begin{align}\begin{split}\label{cofkexpansion}
\hat c(k)\simeq\hat{C}_0 + \hat{C}_2 k^2 +\hat{C}_4 k^4\\
=\hat c(k^*)+(k^2-k^{*2})\hat c^\prime(k^*)
+\frac{1}{2}(k^2-k^{*2})^2\hat c^{\prime\prime}(k^*)\,. 
\end{split}\end{align}
Here, primes denote derivatives with respect to $k^2$. Another way to
determine the coefficients~\cite{Elder:07} is a a fit which
reproduces the isothermal compressibility at $k\rightarrow0$, the
bulk modulus, and the lattice constant of the crystal.

As a subsequent motivation of the suggested fit and also for an
estimate of the phase behavior, we calculated the Helmholtz free
energy of the PFC2 model analytically in the {\it one-mode}
  approximation~\cite{Elder:04}.  Within this approximation, the
two-dimensional density field is assumed to be sinusoidal and
hexagonally symmetric, i.e.,
\begin{equation}
\label{om-phi-eq}
\phi(x,y)\approx A\left[\frac{1}{2}\cos\left( kx\right)-
\cos\left(\frac{\sqrt{3}kx}{2}\right)\cos\left(\frac{ky}{2}\right)
\right]\,,
\end{equation}
with an amplitude $A$ and a nearest neighbor distance of $a=2\pi/k$.
Eq.~\eqref{om-phi-eq} together with eqs.~\eqref{eq:PFC-Fexc} and
\eqref{eq:fid_cpfc} yield the free energy per particle
\begin{equation}\label{fom-eq}
\frac{\mathcal F(A,k)}{N} =
\frac{A^2}{512}\left\{15A^2 - 16A + 96\left[ 1-\hat c(k)\right] 
\right\}\,. 
\end{equation}
Minimization with respect to $k$ and $A$ gives the equilibrium wave
number $k=k^*$ and the equilibrium amplitude
\begin{equation}
\label{ampl-eq}
A^*(\rho)=\left\{\begin{array}{ll}
0\,,&\hat c(k^*;\rho)<c_f\\
\frac{2}{5}\left[\left( 20\,\hat c(k^*;\rho) -19\right)^{1/2} 
+ 1\right],&\hat c(k^*;\rho)>c_f\,,
\end{array}\right.
\end{equation}
where $c_f=43/45=0.956$. The first line corresponds to the stable
fluid and the second to the stable crystal phase, respectively.  For
values of $c_u<\hat c(k^*;\rho)<c_f$, with $c_u=0.95$, the crystalline
density field is metastable, i.e., the free energy,
Eq.~\eqref{fom-eq}, has a local, non-global minimum at a finite
amplitude $A$.  For values $\hat c(k^*;\rho)<c_u$, the crystal is
unstable towards collapse. As the approximate free energy,
Eq.~\eqref{fom-eq}, in the PFC model, is governed by $\hat c(k^*)$, a
proper representation of the latter is on order, and a series
expansion of $\hat c(k)$ about $k^*$ appears natural.

The RY approximation to the DFT predicts a stable crystal for
$\Gamma>\Gamma_f^{\rm DFT}\approx 36.5$ and a metastable crystal for
$\Gamma_u\lesssim\Gamma<\Gamma_f$, with
$\Gamma_u\approx31$~\cite{SvenJPCM}.  The coupling constant at
freezing $\Gamma_f$ corresponds to a maximum value of the correlation
function of $\hat c(k^*;\Gamma_f)=:c_f^{\rm DFT}\approx 0.843$, which
is substantially smaller than the value of $c_f=0.956$, obtained
in the one-mode approximation.
\begin{figure}
\includegraphics[width=8.5cm]{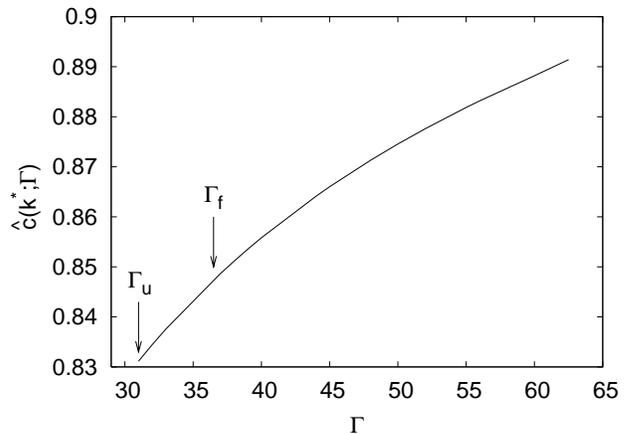}
\begin{center}
  \caption{The maximum of the correlation function $\hat
    c(k^*;\Gamma)$ versus the interaction strength $\Gamma$. The
    arrows at $\Gamma_u$ and $\Gamma_f$ bracket the range of
    metastability of the crystalline density field, obtained from the
    RY approximation to the DFT.\label{Cmax-fig}}
\end{center}
\end{figure}
On the other hand, extrapolation of $\hat c(k^*;\Gamma)$ to a value of
$c_f=0.956$ yields a freezing transition within the one-mode
approximation to the PFC2 model of $\Gamma_{\rm f}\approx 100$, as can
be seen in Fig.~\ref{Cmax-fig}.  In order to obtain a similar
stability regime of the crystalline solution to the DDFT and to the
PFC2 and PFC1 models, the excess free energy of the PFC models is
rescaled by a factor of $f=1.15$, independent of $\Gamma$. The phase
behavior of both PFC models without the constraint on the functional
form of the density field of Eq.~\eqref{om-phi-eq} is similar but
slightly different than in the one-mode approximation; it is discussed
in Subsection~\ref{sec:results}.
\subsection{Non-equilibrium dynamics: Setup description}
\label{subsec:dyn}
In order to measure the propagation front velocities predicted by the DDFT,
Eq.~\eqref{eq:ddft} is numerically solved on a rectangular periodic box of a
fine grid with $\sim64$ grid points per nearest-neighbor distance $a$.  A
finite difference method with variable time step is applied. The convolution
integrals are solved using the method of fast Fourier transform.  The two
versions of the PFC model, eqs.~\eqref{eq:PFC1} and \eqref{eq:PFC2}, are
solved by finite element methods with a variable timestep. The sixth order
partial differential equations are solved semi-implicitly as a system of
second order partial differential equations~\cite{Backofen:07}.

We study the propagation dynamics of the front of a linear array along
the $y$-direction, which is a cutout of a perfect hexagonal crystal
and comprises, at $t=0$, a number of $s$ infinite rows of particles
centered about $x=0$, as can be exemplarily seen from the density map
for $t=0$ in Fig.~\ref{fig:array_evolution} for $\Gamma=60$.  The
number of crystalline rows $s$ is chosen larger than the critical
nucleus to guarantee crystal growth~\cite{SvenPRL} for
$\Gamma>\Gamma_f$ or reasonably large in order to study crystal
shrinkage for $\Gamma<\Gamma_f$\cite{pfc_footnote_s}. The size of the
periodic, rectangular box is therefore chosen integer multiples of the
lattice spacing of the perfectly ordered hexagonal crystal,
\begin{equation}\label{eq:boxsize}
L_x\times L_y= 128(\sqrt{3}/2)a\times a\,.
\end{equation}
The nearest-neighbor distance is fixed to its ideal value
\begin{equation}\label{eq:latticeconst}
a=(2/\sqrt{3})^{1/2}\rho^{-1/2}\,,
\end{equation}
which is very close to the equilibrium value of the one-mode
approximation to the PFC2 model, $a=2\pi/k^*$and to the equilibrium
value of the RY approximation to the DFT for all values of
$\Gamma$~\cite{SvenJPCM}. The initial density field is given by
\begin{equation}\label{eq:rho_r}
  \rho({\bf r},t=0)= \left[1-h(|x|-R)\right]\rho_{\rm c}({\bf r}) 
  + h(|x|-R)\rho\,.
\end{equation}
Here, $h(x)$ is a smoothed approximation to the Heaviside step
function. $\rho_{\rm c}({\bf r})$ is the infinite, stable or
metastable, crystalline density field with constrained lattice
constant $a$ and with the $[11]$-orientation parallel to the $x$-axis,
which is symmetric about $x=0$.  The size of the initial nucleus is
defined by $R=(\sqrt{3}/4)sa$, with $s$ the number of crystalline rows
parallel to the $y$-axis. $h(|x|-R)$ is therefore chosen to cut
through valleys of the density field guaranteeing the overall particle
number to be fixed, i.e., $V^{-1}\int_V{\mathrm d} {\bf r}\rho({\bf
  r},t=0)=\rho$. For the number of crystalline rows for the different
coupling constants, see~\cite{pfc_footnote_s}.

The initial density field $\rho({\bf r},t=0)$ can be thought of as an
equilibrium density field with an appropriately chosen, though
somewhat artificial, external potential $V({\bf r})$~\cite{Oxtoby:91}.
An experimentally more feasible and thus more realistic setup has been
suggested in reference~\cite{SvenPRL}: The array of tagged particles
was first, i.e., for times $t<0$, held fixed in a thermodynamically
stable, equilibrated fluid of density $\rho$ at a coupling constant
$\Gamma<\Gamma_f$ well below freezing. For the equilibration of the
fluid, Eq.~(\ref{eq:ddft}) was numerically solved fixing the tagged
particles by deep parabolic external potentials at the tagged particle
positions---in an experiment this could be achieved by using optical
tweezers~\cite{Koppl:06}.  At time $t=0$ the external pinning
potential was turned off and, at the same time, the system was
instantaneously quenched to a a higher coupling constant $\Gamma$.
Experimentally, the instantaneous quench is easily achieved by
increasing the homogeneous external magnetic field ${\bf B}$. Both
protocols lead to different short-time dynamics [$t\lesssim(\rho
D)^{-1}$] due to the initial difference in the density field close to
the incipient cluster.  However, for longer times the density fields
are indistinguishable (data not shown) thus leading to the same
propagation front velocities which are of interest in this work.

\subsection{Non-equilibrium dynamics: Results}
\label{sec:results}
\begin{figure}
  \includegraphics[width=8.5cm]{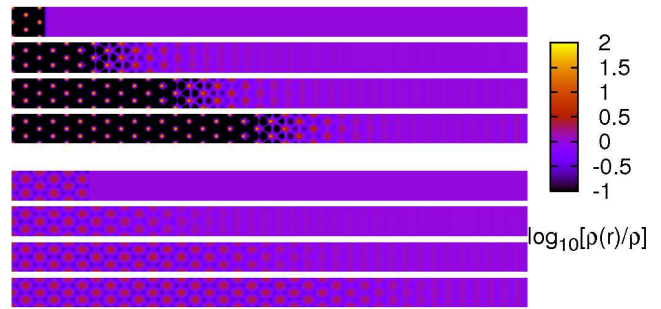}
  \caption{(Color online) Snapshots of of the dimensionless,
    logarithmic density field $\log_{10}[\rho({\bf r},t)/\rho]$ of a
    linear nucleus of initially $s=5$ (DDFT; top panel) or $s=11$
    (PFC1 model; bottom panel) infinite rows of hexagonally
    crystalline particles at coupling constant $\Gamma=60$, as
    obtained from the DDFT (top panel) and the PFC1 model (bottom
    panel). The upper and the lower four maps show the density fields
    each at times $t/\tau_B=0, 0.5, 1, 1.5$ (from top to bottom). For
    better visibility and exploiting the symmetry of the density
    field, the images display twice the right half of the system's
    central region of dimensions $35\sqrt{3}/2a\times 2a$.  }
  \label{fig:array_evolution}
\end{figure}

\begin{figure}
   \includegraphics[width=8.5cm]{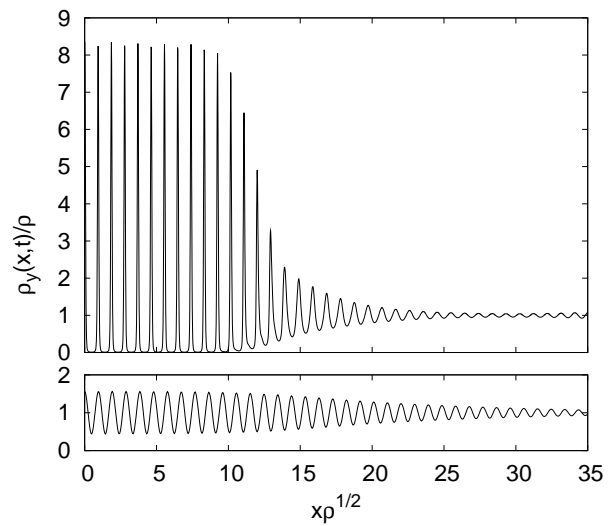}
   \caption{The $y$-average density profile $\rho_y(x,t=\tau_B)$
     obtained from the DDFT (top panel) and the PFC1 model (bottom
     panel) for the same coupling constant $\Gamma=60$ and the same
     initial conditions as in Fig.~\ref{fig:array_evolution} at the
     time $t/\tau_B=1$.}
  \label{fig:array_rhoryavg}
\end{figure}

In Fig.~\ref{fig:array_evolution}, we display the density field
$\rho({\bf r},t)$ as obtained from the DDFT and from the rescaled PFC2
model for $\Gamma=60$ at four different times $t/\tau_B=0, 0.5, 1,
1.5$, where we chose the Brownian time scale $\tau_B=(\rho D)^{-1}$ as
the unit time scale.
Since the density field is symmetric with respect to the $x$-axis we
concentrate on the region $x>0$ in the following.
Fig.~\ref{fig:array_rhoryavg} displays the corresponding $y$-averaged
density profile,
\begin{equation}
  \label{eq:2}
  \rho_y(x,t)\equiv L_y^{-1} \int \mathrm d
  y\,\rho({\bf r},t) \,,
\end{equation}
at a short time $t=\tau_B$, which is large enough for $\rho_y({\bf
  r},t)$ to be insensitive on the details of the same field at time
$t=0$.
Density fields obtained within the rescaled PFC1 model are
qualitatively very similar to the ones of the PFC2 model; they are
therefore not displayed in the present paper.

It can be ascertained from Fig.~\ref{fig:array_evolution} that after
short time the propagating crystal fronts approximately establish
steady states, which eventually change on larger time scales
$t\gg\tau_B$~\cite{pfc_footnote_steadystate}.  As can also be seen
from Figs.~\ref{fig:array_evolution} and \ref{fig:array_rhoryavg}, the
crystalline density fields behind the crystal fronts, which are close
to equilibrium in all three models, are modulated much stronger about
the average density $\rho$ in the DDFT than in the two rescaled PFC
models.  This difference goes along with an almost sinusoidal density
field in the PFC models, approximately equal to the field assumed in
the one-mode approximation, Eq.~\eqref{om-phi-eq}, which is due to the
truncation of the expansion of $F_{\rm ex}[\rho]$ in $k^2$
[Eq.~\eqref{eq:PFC-Fexc}]; this density field is in contrast to the
one of the DDFT, which is approximately given by a superposition of
Gaussians centered about the lattice vectors of the hexagonal lattice.
Another qualitative difference between the two models concerns the
crystal-melt interface: the width of the crystal front obtained within
the DDFT is substantially smaller than in the PFC models; whereas the
former is of the order of $\Delta x\sim5\rho^{-1/2}$, the latter
approximately amounts $\Delta x\sim25\rho^{-1/2}$. In both models, the
widths are relatively insensitive towards changes in coupling constant
$\Gamma$ (data not shown).

\begin{figure}
\includegraphics[width=8.5cm]{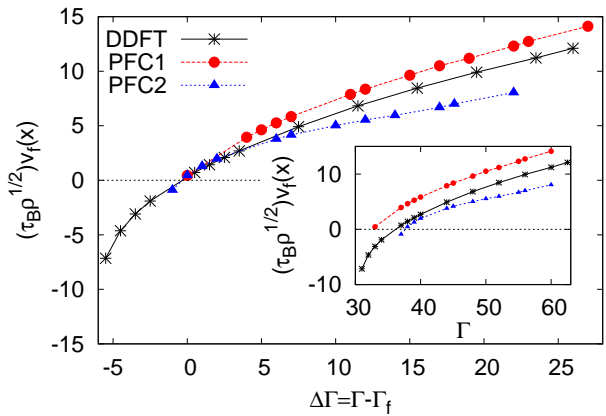}
\caption{(Color online) Propagation front velocities of a linear
  crystal front in the $[11]$-direction, $v_f(\Gamma)$, measured
  shortly after the quench (see text), as a function of relative
  coupling constant $\Delta\Gamma=\Gamma-\Gamma_f$, with $\Gamma_f$
  the respective coupling constant of freezing, obtained within the
  DDFT (stars), the PFC1 model (circles), and the PFC2 model
  (triangles). Lines are guides to the eye. Inset: The same velocities
  as a function of $\Gamma$.}
\label{fig:velocity}
\end{figure}

In order to quantify the crystal front propagation, the position of
the diffuse front, $x_f(t)$, is extracted as the maximum $x$-position
at which the density field exceeds the value $2\rho$, in the DDFT, and
as the inflection point of the envelope function to $\rho_y(x,t)$ in
the PFC model, respectively (data not shown).  Within the DDFT, the
front position is thus situated at a position, which is at the same
time slightly larger than the one in the PFC models.  However, this
does not affect the crystal front velocity $v_f(\Gamma)=\partial
x_f(t)/\partial t$, which was measured as a function of coupling
constant $\Gamma$ for the three different models under study at a
short distance from the incipient front,
$x_f(t=0)+15\rho^{-1/2}<x_f(t)<x_f(t=0)+20\rho^{-1/2}$ (see
Fig.~\ref{fig:velocity}).
The front velocity at long times is not considered in this paper.
Comparing the three different curves of Fig.~\ref{fig:velocity}, the
following observations are made:

(i) In all three models, $v_f(\Gamma)$ increases monotonically from
$0$ with increasing difference in coupling constant
$\Delta\Gamma\equiv\Gamma-\Gamma_f$, where the freezing constant is
$\Gamma_f=36.5$ within the DFT~\cite{SvenPhilMag}, $\Gamma_f\approx33$
in the rescaled PFC1 model, and $\Gamma_f\approx38$ within the
rescaled PFC2 model. The three coupling constants do not agree because
the rescaling of the excess free energy was chosen to bring only the
freezing constants of the more approximate one-mode approximation to
the PFC2 model into agreement with the freezing constant of the DFT.
The same scaling is then used for the PFC1 and PFC2 models without the
constraint of a sinusoidal density field, Eq.~\eqref{om-phi-eq}. The
difference between the respective values within the PFC1 and the PFC2
models is due to the difference in their respective ideal free
energies $F_{\rm id}[\rho]$, Eqs.~\eqref{eq:fid} and
\eqref{eq:fid_cpfc}.  Moreover, as the ideal free energy functionals
of the DFT and the PFC1 model are equal, the freezing constant of the
rescaled PFC1 model is lower than the one of the rescaled PFC2 model,
which is reminiscent of the smaller distance of the respective
freezing constants within the non-rescaled models (obtained from an
extrapolation of $\hat c(k)$; data not shown).

(ii) For negative $\Delta\Gamma$, with $\Gamma$ still within the
metastability regime, $\Gamma_u<\Gamma<\Gamma_f$, the propagation
front velocity is negative and a retraction of an almost steady-state
crystal front is observed. 

(iii) For $\Delta\Gamma>0$, the front velocities obtained within the
two rescaled PFC models bracket the respective values obtained within
the DDFT, the PFC1 model being slightly closer to the DDFT as the PFC2
model.  All three models yield similar front velocities, although the
corresponding steady-state density fields are quite different (see
Fig.~\ref{fig:array_rhoryavg}). However, the non-monotonicity of the
propagation front velocities $v_f(\Delta\Gamma)$ for the same value of
$\Delta\Gamma$ with increasing degree of approximation (from the DDFT
via the PFC1 model to the PFC2 model) points to a cancellation of
errors in the approximate PFC models. Still, as would have been
expected, the velocities obtained from the less approximate PFC1 model
are closer to the results of the DDFT than those of the more
approximate PFC2 model.
%
\section{Conclusions}
\label{sec:conclusions}
In conclusion, we have derived the phase field crystal (PFC) model
from microscopic dynamical density functional theory (DDFT)
appropriate for colloidal dispersions, which are governed by
completely overdamped Brownian dynamics. The ordinary phase field
crystal model (called PFC2 model) arises from a constant mobility
approximation {\it and\/} an expansion of the ideal gas entropy in
terms of density.  Both approximations can be avoided yielding a
variant of the phase field crystal model (called PFC1 model), which
requires the same computational effort as the PFC2 model.

Comparing the two phase field crystal models to the full solution of
the dynamical density functional theory, agreement could only be
obtained by an empirical scaling factor in the free energy.  On the
one hand, this implies that phase field crystal models have to be used
with care and lack full ab initio precision. If a suitable scaling is
accepted, there is good overall agreement in the growth velocity of a
crystalline front, where the PFC1 model yields slightly better
agreement than the PFC2 model. The corresponding density fields,
however, differ vastly in terms of the sharpness of the crystalline
peaks.  We therefore conclude that the phase field crystal model gives
a qualitative reliable description of the trends in crystal growth
processes but lacks high precision.

Future work should focus, first, on three-dimensional crystalline
fronts and on the dynamics and annealing of crystalline defects, where
both DDFT and PFC models should be compared as well. Second, the
behavior of the setup at intermediate and long times deserves more
detailed study.  Finally, full Brownian dynamics computer simulations
\cite{SvenPRL} are needed to obtain reference data for a test of the
underlying adiabatic approximation in the DDFT.
\begin{acknowledgments}
  We thank R.~Blaak and C.\ N.\ Likos for helpful discussions. This
  work has been supported by the DFG through the DFG priority program
  SPP 1296.
\end{acknowledgments}



\begin{thebibliography}{63}
\expandafter\ifx\csname natexlab\endcsname\relax\def\natexlab#1{#1}\fi
\expandafter\ifx\csname bibnamefont\endcsname\relax
  \def\bibnamefont#1{#1}\fi
\expandafter\ifx\csname bibfnamefont\endcsname\relax
  \def\bibfnamefont#1{#1}\fi
\expandafter\ifx\csname citenamefont\endcsname\relax
  \def\citenamefont#1{#1}\fi
\expandafter\ifx\csname url\endcsname\relax
  \def\url#1{\texttt{#1}}\fi
\expandafter\ifx\csname urlprefix\endcsname\relax\def\urlprefix{URL }\fi
\providecommand{\bibinfo}[2]{#2}
\providecommand{\eprint}[2][]{\url{#2}}

\bibitem[{\citenamefont{Wiener}(2004)}]{Wiener:04}
\bibinfo{author}{\bibfnamefont{M.~C.} \bibnamefont{Wiener}},
  \bibinfo{journal}{Methods} \textbf{\bibinfo{volume}{34}},
  \bibinfo{pages}{364} (\bibinfo{year}{2004}).

\bibitem[{\citenamefont{Snell and Helliwell}(2005)}]{Snell:05}
\bibinfo{author}{\bibfnamefont{E.~H.} \bibnamefont{Snell}} \bibnamefont{and}
  \bibinfo{author}{\bibfnamefont{J.~R.} \bibnamefont{Helliwell}},
  \bibinfo{journal}{Rep.~Prog.~Phys.} \textbf{\bibinfo{volume}{68}},
  \bibinfo{pages}{799} (\bibinfo{year}{2005}).

\bibitem[{\citenamefont{Dziomkina and Vancso}(2005)}]{Dziomkina:05}
\bibinfo{author}{\bibfnamefont{N.~V.} \bibnamefont{Dziomkina}}
  \bibnamefont{and} \bibinfo{author}{\bibfnamefont{G.~J.}
  \bibnamefont{Vancso}}, \bibinfo{journal}{Soft Matter}
  \textbf{\bibinfo{volume}{1}}, \bibinfo{pages}{265} (\bibinfo{year}{2005}).

\bibitem[{\citenamefont{Emmerich et~al.}(2007)\citenamefont{Emmerich, Binder,
  and Nestler}}]{Emmerich:07}
\bibinfo{author}{\bibfnamefont{H.}~\bibnamefont{Emmerich}},
  \bibinfo{author}{\bibfnamefont{K.}~\bibnamefont{Binder}}, \bibnamefont{and}
  \bibinfo{author}{\bibfnamefont{B.}~\bibnamefont{Nestler}},
  \bibinfo{journal}{Philos.~Mag.~Lett.} \textbf{\bibinfo{volume}{87}},
  \bibinfo{pages}{791} (\bibinfo{year}{2007}).

\bibitem[{\citenamefont{Emmerich}(2008)}]{Emmerich:08}
\bibinfo{author}{\bibfnamefont{H.}~\bibnamefont{Emmerich}},
  \bibinfo{journal}{Adv. Phys.} \textbf{\bibinfo{volume}{57}},
  \bibinfo{pages}{1} (\bibinfo{year}{2008}).

\bibitem[{\citenamefont{Elder et~al.}(2002)\citenamefont{Elder, Katakowski,
  Haataja, and Grant}}]{Elder:02}
\bibinfo{author}{\bibfnamefont{K.~R.} \bibnamefont{Elder}},
  \bibinfo{author}{\bibfnamefont{M.}~\bibnamefont{Katakowski}},
  \bibinfo{author}{\bibfnamefont{M.}~\bibnamefont{Haataja}}, \bibnamefont{and}
  \bibinfo{author}{\bibfnamefont{M.}~\bibnamefont{Grant}},
  \bibinfo{journal}{Phys.~Rev.~Lett.} \textbf{\bibinfo{volume}{88}},
  \bibinfo{pages}{245701} (\bibinfo{year}{2002}).

\bibitem[{\citenamefont{Athreya et~al.}(2007)\citenamefont{Athreya, Goldenfeld,
  Dantzig, Greenwood, and Provatas}}]{Athreya:07}
\bibinfo{author}{\bibfnamefont{B.~P.} \bibnamefont{Athreya}},
  \bibinfo{author}{\bibfnamefont{N.}~\bibnamefont{Goldenfeld}},
  \bibinfo{author}{\bibfnamefont{J.~A.} \bibnamefont{Dantzig}},
  \bibinfo{author}{\bibfnamefont{M.}~\bibnamefont{Greenwood}},
  \bibnamefont{and} \bibinfo{author}{\bibfnamefont{N.}~\bibnamefont{Provatas}},
  \bibinfo{journal}{Phys.~Rev.~E} \textbf{\bibinfo{volume}{76}},
  \bibinfo{pages}{056706} (\bibinfo{year}{2007}).

\bibitem[{\citenamefont{Wu and Karma}(2007)}]{Wu:07}
\bibinfo{author}{\bibfnamefont{K.~A.} \bibnamefont{Wu}} \bibnamefont{and}
  \bibinfo{author}{\bibfnamefont{A.}~\bibnamefont{Karma}},
  \bibinfo{journal}{Phys.~Rev.~B} \textbf{\bibinfo{volume}{76}},
  \bibinfo{pages}{184107} (\bibinfo{year}{2007}).

\bibitem[{\citenamefont{Goldenfeld et~al.}(2005)\citenamefont{Goldenfeld,
  Athreya, and Dantzig}}]{Goldenfeld:05}
\bibinfo{author}{\bibfnamefont{N.}~\bibnamefont{Goldenfeld}},
  \bibinfo{author}{\bibfnamefont{B.~P.} \bibnamefont{Athreya}},
  \bibnamefont{and} \bibinfo{author}{\bibfnamefont{J.~A.}
  \bibnamefont{Dantzig}}, \bibinfo{journal}{Phys.~Rev.~E}
  \textbf{\bibinfo{volume}{72}}, \bibinfo{pages}{020601(R)}
  (\bibinfo{year}{2005}).

\bibitem[{\citenamefont{Backofen et~al.}(2007)\citenamefont{Backofen, Ratz, and
  Voigt}}]{Backofen:07}
\bibinfo{author}{\bibfnamefont{R.}~\bibnamefont{Backofen}},
  \bibinfo{author}{\bibfnamefont{A.}~\bibnamefont{Ratz}}, \bibnamefont{and}
  \bibinfo{author}{\bibfnamefont{A.}~\bibnamefont{Voigt}},
  \bibinfo{journal}{Philos.~Mag.~Lett.} \textbf{\bibinfo{volume}{87}},
  \bibinfo{pages}{813} (\bibinfo{year}{2007}).

\bibitem[{\citenamefont{Achim et~al.}(2006)\citenamefont{Achim, Karttunen,
  Elder, Granato, Ala-Nissila, and Ying}}]{Achim:06}
\bibinfo{author}{\bibfnamefont{C.~V.} \bibnamefont{Achim}},
  \bibinfo{author}{\bibfnamefont{M.}~\bibnamefont{Karttunen}},
  \bibinfo{author}{\bibfnamefont{K.~R.} \bibnamefont{Elder}},
  \bibinfo{author}{\bibfnamefont{E.}~\bibnamefont{Granato}},
  \bibinfo{author}{\bibfnamefont{T.}~\bibnamefont{Ala-Nissila}},
  \bibnamefont{and} \bibinfo{author}{\bibfnamefont{S.~C.} \bibnamefont{Ying}},
  \bibinfo{journal}{Phys.~Rev.~E} \textbf{\bibinfo{volume}{74}},
  \bibinfo{pages}{021104} (\bibinfo{year}{2006}).

\bibitem[{\citenamefont{Berry et~al.}(2006)\citenamefont{Berry, Grant, and
  Elder}}]{Berry:06}
\bibinfo{author}{\bibfnamefont{J.}~\bibnamefont{Berry}},
  \bibinfo{author}{\bibfnamefont{M.}~\bibnamefont{Grant}}, \bibnamefont{and}
  \bibinfo{author}{\bibfnamefont{K.~R.} \bibnamefont{Elder}},
  \bibinfo{journal}{Phys.~Rev.~E} \textbf{\bibinfo{volume}{73}},
  \bibinfo{pages}{031609} (\bibinfo{year}{2006}).

\bibitem[{\citenamefont{Oxtoby}(1991)}]{Oxtoby:91}
\bibinfo{author}{\bibfnamefont{D.~W.} \bibnamefont{Oxtoby}}, in
  \emph{\bibinfo{booktitle}{Liquids, Freezing and the Glass Transition}}
  (\bibinfo{publisher}{North Holland}, \bibinfo{address}{Amsterdam},
  \bibinfo{year}{1991}), vol. \bibinfo{volume}{Session LI (1989)} of
  \emph{\bibinfo{series}{Les Houches Summer Schools of Theoretical Physics
  Session LI (1989)}}, p. \bibinfo{pages}{147}.

\bibitem[{\citenamefont{Singh}(1991)}]{Singh:91}
\bibinfo{author}{\bibfnamefont{Y.}~\bibnamefont{Singh}},
  \bibinfo{journal}{Phys.~Rep.} \textbf{\bibinfo{volume}{207}},
  \bibinfo{pages}{351} (\bibinfo{year}{1991}).

\bibitem[{\citenamefont{L{\"o}wen}(1994)}]{Loewen:94}
\bibinfo{author}{\bibfnamefont{H.}~\bibnamefont{L{\"o}wen}},
  \bibinfo{journal}{Phys.~Rep.} \textbf{\bibinfo{volume}{237}},
  \bibinfo{pages}{249} (\bibinfo{year}{1994}).

\bibitem[{\citenamefont{L{\"o}wen}(2002)}]{Loewen:02}
\bibinfo{author}{\bibfnamefont{H.}~\bibnamefont{L{\"o}wen}},
  \bibinfo{journal}{J.~Phys.:~Condens.~Matter} \textbf{\bibinfo{volume}{14}},
  \bibinfo{pages}{11897} (\bibinfo{year}{2002}).

\bibitem[{\citenamefont{Likos}(2001)}]{Likos:01}
\bibinfo{author}{\bibfnamefont{C.~N.} \bibnamefont{Likos}},
  \bibinfo{journal}{Phys. Rep.} \textbf{\bibinfo{volume}{348}},
  \bibinfo{pages}{267} (\bibinfo{year}{2001}).

\bibitem[{\citenamefont{Rex et~al.}(2005)\citenamefont{Rex, L{\"o}wen, and
  Likos}}]{Rex1}
\bibinfo{author}{\bibfnamefont{M.}~\bibnamefont{Rex}},
  \bibinfo{author}{\bibfnamefont{H.}~\bibnamefont{L{\"o}wen}},
  \bibnamefont{and} \bibinfo{author}{\bibfnamefont{C.~N.} \bibnamefont{Likos}},
  \bibinfo{journal}{Phys.~Rev.~E} \textbf{\bibinfo{volume}{72}},
  \bibinfo{pages}{021404} (\bibinfo{year}{2005}).

\bibitem[{\citenamefont{Rex et~al.}(2006)\citenamefont{Rex, Likos, L{\"o}wen,
  and Dzubiella}}]{Rex:06}
\bibinfo{author}{\bibfnamefont{M.}~\bibnamefont{Rex}},
  \bibinfo{author}{\bibfnamefont{C.~N.} \bibnamefont{Likos}},
  \bibinfo{author}{\bibfnamefont{H.}~\bibnamefont{L{\"o}wen}},
  \bibnamefont{and}
  \bibinfo{author}{\bibfnamefont{J.}~\bibnamefont{Dzubiella}},
  \bibinfo{journal}{Mol.~Phys.} \textbf{\bibinfo{volume}{104}},
  \bibinfo{pages}{527} (\bibinfo{year}{2006}).

\bibitem[{\citenamefont{Archer and Evans}(2004)}]{Archer:04}
\bibinfo{author}{\bibfnamefont{A.~J.} \bibnamefont{Archer}} \bibnamefont{and}
  \bibinfo{author}{\bibfnamefont{R.}~\bibnamefont{Evans}},
  \bibinfo{journal}{J.~Chem.~Phys.} \textbf{\bibinfo{volume}{121}},
  \bibinfo{pages}{4246} (\bibinfo{year}{2004}).

\bibitem[{\citenamefont{Marconi and Tarazona}(1999)}]{Marconi:99}
\bibinfo{author}{\bibfnamefont{U.~M.~B.} \bibnamefont{Marconi}}
  \bibnamefont{and} \bibinfo{author}{\bibfnamefont{P.}~\bibnamefont{Tarazona}},
  \bibinfo{journal}{J.~Chem.~Phys.} \textbf{\bibinfo{volume}{110}},
  \bibinfo{pages}{8032} (\bibinfo{year}{1999}).

\bibitem[{\citenamefont{Dzubiella and Likos}(2003)}]{Dzubiella:03}
\bibinfo{author}{\bibfnamefont{J.}~\bibnamefont{Dzubiella}} \bibnamefont{and}
  \bibinfo{author}{\bibfnamefont{C.~N.} \bibnamefont{Likos}},
  \bibinfo{journal}{J.~Phys.:~Condens.~Matter} \textbf{\bibinfo{volume}{15}},
  \bibinfo{pages}{L147} (\bibinfo{year}{2003}).

\bibitem[{\citenamefont{L{\"o}wen et~al.}(2007)\citenamefont{L{\"o}wen, Likos,
  Assoud, Blaak, and van Teeffelen}}]{SvenPhilMag}
\bibinfo{author}{\bibfnamefont{H.}~\bibnamefont{L{\"o}wen}},
  \bibinfo{author}{\bibfnamefont{C.~N.} \bibnamefont{Likos}},
  \bibinfo{author}{\bibfnamefont{L.}~\bibnamefont{Assoud}},
  \bibinfo{author}{\bibfnamefont{R.}~\bibnamefont{Blaak}}, \bibnamefont{and}
  \bibinfo{author}{\bibfnamefont{S.}~\bibnamefont{van Teeffelen}},
  \bibinfo{journal}{Philos.~Mag.~Lett.} \textbf{\bibinfo{volume}{87}},
  \bibinfo{pages}{847} (\bibinfo{year}{2007}).

\bibitem[{\citenamefont{van Teeffelen
  et~al.}(2008{\natexlab{a}})\citenamefont{van Teeffelen, Likos, and
  L{\"o}wen}}]{SvenPRL}
\bibinfo{author}{\bibfnamefont{S.}~\bibnamefont{van Teeffelen}},
  \bibinfo{author}{\bibfnamefont{C.~N.} \bibnamefont{Likos}}, \bibnamefont{and}
  \bibinfo{author}{\bibfnamefont{H.}~\bibnamefont{L{\"o}wen}},
  \bibinfo{journal}{Phys.~Rev.~Lett.} \textbf{\bibinfo{volume}{100}},
  \bibinfo{pages}{108302} (\bibinfo{year}{2008}{\natexlab{a}}).

\bibitem[{\citenamefont{Elder et~al.}(2007)\citenamefont{Elder, Provatas,
  Berry, Stefanovic, and Grant}}]{Elder:07}
\bibinfo{author}{\bibfnamefont{K.~R.} \bibnamefont{Elder}},
  \bibinfo{author}{\bibfnamefont{N.}~\bibnamefont{Provatas}},
  \bibinfo{author}{\bibfnamefont{J.}~\bibnamefont{Berry}},
  \bibinfo{author}{\bibfnamefont{P.}~\bibnamefont{Stefanovic}},
  \bibnamefont{and} \bibinfo{author}{\bibfnamefont{M.}~\bibnamefont{Grant}},
  \bibinfo{journal}{Phys.~Rev.~B} \textbf{\bibinfo{volume}{75}},
  \bibinfo{pages}{064107} (\bibinfo{year}{2007}).

\bibitem[{Low()}]{Lowencrystalmult}
\bibinfo{note}{H.\ L{\"o}wen, T.\ Beier, H.\ Wagner, Europhys. Lett. {\bf 9},
  791 (1989); Z.\ Phys.\ B {\bf 79}, 109 (1990).}

\bibitem[{\citenamefont{Lutsko}(2006)}]{Lutsko:06}
\bibinfo{author}{\bibfnamefont{J.~F.} \bibnamefont{Lutsko}},
  \bibinfo{journal}{Physica A} \textbf{\bibinfo{volume}{366}},
  \bibinfo{pages}{229} (\bibinfo{year}{2006}).

\bibitem[{\citenamefont{Risken}(1989)}]{Risken:89}
\bibinfo{author}{\bibfnamefont{H.}~\bibnamefont{Risken}},
  \emph{\bibinfo{title}{The {F}okker-{P}lanck Equation, Methods of Solution and
  Applications}} (\bibinfo{publisher}{Springer}, \bibinfo{address}{Berlin},
  \bibinfo{year}{1989}), \bibinfo{edition}{2nd} ed.

\bibitem[{\citenamefont{von Smoluchowski}(1916)}]{Smoluchowski:15}
\bibinfo{author}{\bibfnamefont{M.}~\bibnamefont{von Smoluchowski}},
  \bibinfo{journal}{Ann. Phys. (Leipzig)} \textbf{\bibinfo{volume}{353}},
  \bibinfo{pages}{1103} (\bibinfo{year}{1916}).

\bibitem[{\citenamefont{Zahn et~al.}(1997)\citenamefont{Zahn, M\'endez-Alcaraz,
  and Maret}}]{Zahn:97}
\bibinfo{author}{\bibfnamefont{K.}~\bibnamefont{Zahn}},
  \bibinfo{author}{\bibfnamefont{J.~M.} \bibnamefont{M\'endez-Alcaraz}},
  \bibnamefont{and} \bibinfo{author}{\bibfnamefont{G.}~\bibnamefont{Maret}},
  \bibinfo{journal}{Phys.~Rev.~Lett.} \textbf{\bibinfo{volume}{79}},
  \bibinfo{pages}{175} (\bibinfo{year}{1997}).

\bibitem[{\citenamefont{Haghgooie and Doyle}(2005)}]{Haghgooie:05}
\bibinfo{author}{\bibfnamefont{R.}~\bibnamefont{Haghgooie}} \bibnamefont{and}
  \bibinfo{author}{\bibfnamefont{P.~S.} \bibnamefont{Doyle}},
  \bibinfo{journal}{Phys.~Rev.~E} \textbf{\bibinfo{volume}{72}},
  \bibinfo{pages}{011405} (\bibinfo{year}{2005}).

\bibitem[{\citenamefont{Ramakrishnan and Yussouff}(1979)}]{Ramakrishnan:79}
\bibinfo{author}{\bibfnamefont{T.~V.} \bibnamefont{Ramakrishnan}}
  \bibnamefont{and} \bibinfo{author}{\bibfnamefont{M.}~\bibnamefont{Yussouff}},
  \bibinfo{journal}{Phys.~Rev.~B} \textbf{\bibinfo{volume}{19}},
  \bibinfo{pages}{2775} (\bibinfo{year}{1979}).

\bibitem[{\citenamefont{Dean}(1996)}]{Dean:96}
\bibinfo{author}{\bibfnamefont{D.~S.} \bibnamefont{Dean}},
  \bibinfo{journal}{J.~Phys.~A.: Math. Gen.} \textbf{\bibinfo{volume}{29}},
  \bibinfo{pages}{L613} (\bibinfo{year}{1996}).

\bibitem[{\citenamefont{Enskog}(1922) [English translation in Brush, S. G.
  (editor), {\it Kinetic Theory}, vol. 3, Pergamon Press, London, New York,
  1972, (p. 226)])}]{Enskog:22}
\bibinfo{author}{\bibfnamefont{D.}~\bibnamefont{Enskog}}, \bibinfo{journal}{K.
  Sven. Vetenskapsakad. Handl.} \textbf{\bibinfo{volume}{63}},
  \bibinfo{pages}{4} (\bibinfo{year}{1922) [English translation in Brush, S. G.
  (editor), {\it Kinetic Theory}, vol. 3, Pergamon Press, London, New York,
  1972, (p. 226)]}).

\bibitem[{\citenamefont{Evans}(1979)}]{Evans:79}
\bibinfo{author}{\bibfnamefont{R.}~\bibnamefont{Evans}}, \bibinfo{journal}{Adv.
  Phys.} \textbf{\bibinfo{volume}{28}}, \bibinfo{pages}{143}
  (\bibinfo{year}{1979}).

\bibitem[{\citenamefont{Marconi and Tarazona}(2000)}]{Marconi:00}
\bibinfo{author}{\bibfnamefont{U.~M.~B.} \bibnamefont{Marconi}}
  \bibnamefont{and} \bibinfo{author}{\bibfnamefont{P.}~\bibnamefont{Tarazona}},
  \bibinfo{journal}{J.~Phys.:~Condens.~Matter} \textbf{\bibinfo{volume}{12}},
  \bibinfo{pages}{A413} (\bibinfo{year}{2000}).

\bibitem[{\citenamefont{Archer and Rauscher}(2004)}]{Rauscher}
\bibinfo{author}{\bibfnamefont{A.}~\bibnamefont{Archer}} \bibnamefont{and}
  \bibinfo{author}{\bibfnamefont{M.}~\bibnamefont{Rauscher}},
  \bibinfo{journal}{J.~Phys.~A.: Math. Gen.} \textbf{\bibinfo{volume}{37}},
  \bibinfo{pages}{9325} (\bibinfo{year}{2004}).

\bibitem[{\citenamefont{Dieterich et~al.}(1990)\citenamefont{Dieterich, Frisch,
  and Majhofer}}]{Dieterich:90}
\bibinfo{author}{\bibfnamefont{W.}~\bibnamefont{Dieterich}},
  \bibinfo{author}{\bibfnamefont{H.~L.} \bibnamefont{Frisch}},
  \bibnamefont{and} \bibinfo{author}{\bibfnamefont{A.}~\bibnamefont{Majhofer}},
  \bibinfo{journal}{Z.~Phys.~B} \textbf{\bibinfo{volume}{78}},
  \bibinfo{pages}{317} (\bibinfo{year}{1990}).

\bibitem[{\citenamefont{Haymet and Oxtoby}(1981)}]{Haymet:81}
\bibinfo{author}{\bibfnamefont{A.~D.~J.} \bibnamefont{Haymet}}
  \bibnamefont{and} \bibinfo{author}{\bibfnamefont{D.~W.}
  \bibnamefont{Oxtoby}}, \bibinfo{journal}{J.~Chem.~Phys.}
  \textbf{\bibinfo{volume}{74}}, \bibinfo{pages}{2559} (\bibinfo{year}{1981}).

\bibitem[{\citenamefont{Munakata}(1977{\natexlab{a}})}]{Munakata:77}
\bibinfo{author}{\bibfnamefont{T.}~\bibnamefont{Munakata}},
  \bibinfo{journal}{J.~Phys.~Soc.~Jap.} \textbf{\bibinfo{volume}{43}},
  \bibinfo{pages}{1762} (\bibinfo{year}{1977}{\natexlab{a}}).

\bibitem[{\citenamefont{Munakata}(1977{\natexlab{b}})}]{Munakata:77b}
\bibinfo{author}{\bibfnamefont{T.}~\bibnamefont{Munakata}},
  \bibinfo{journal}{J.~Phys.~Soc.~Jap.} \textbf{\bibinfo{volume}{43}},
  \bibinfo{pages}{1723} (\bibinfo{year}{1977}{\natexlab{b}}).

\bibitem[{\citenamefont{Calef and Wolynes}(1983)}]{Calef:83}
\bibinfo{author}{\bibfnamefont{D.~F.} \bibnamefont{Calef}} \bibnamefont{and}
  \bibinfo{author}{\bibfnamefont{P.~G.} \bibnamefont{Wolynes}},
  \bibinfo{journal}{J.~Chem.~Phys.} \textbf{\bibinfo{volume}{78}},
  \bibinfo{pages}{4145} (\bibinfo{year}{1983}).

\bibitem[{\citenamefont{Chandra and Bagchi}(1989)}]{Chandra:89}
\bibinfo{author}{\bibfnamefont{A.}~\bibnamefont{Chandra}} \bibnamefont{and}
  \bibinfo{author}{\bibfnamefont{B.}~\bibnamefont{Bagchi}},
  \bibinfo{journal}{J.~Chem.~Phys.} \textbf{\bibinfo{volume}{91}},
  \bibinfo{pages}{1829} (\bibinfo{year}{1989}).

\bibitem[{\citenamefont{Chandra and Bagchi}(1990)}]{Chandra:90}
\bibinfo{author}{\bibfnamefont{A.}~\bibnamefont{Chandra}} \bibnamefont{and}
  \bibinfo{author}{\bibfnamefont{B.}~\bibnamefont{Bagchi}},
  \bibinfo{journal}{Physica A} \textbf{\bibinfo{volume}{169}},
  \bibinfo{pages}{246} (\bibinfo{year}{1990}).

\bibitem[{\citenamefont{R{\'e}sibois and
  DeLeener}(1977)}]{resibois-deleener:77}
\bibinfo{author}{\bibfnamefont{P.}~\bibnamefont{R{\'e}sibois}}
  \bibnamefont{and} \bibinfo{author}{\bibfnamefont{M.}~\bibnamefont{DeLeener}},
  \emph{\bibinfo{title}{Classical Kinetic Theory of Fluids}}
  (\bibinfo{publisher}{John Wiley}, \bibinfo{address}{New York},
  \bibinfo{year}{1977}).

\bibitem[{\citenamefont{Kirkpatrick}(1989)}]{Kirkpatrick:89}
\bibinfo{author}{\bibfnamefont{T.~R.} \bibnamefont{Kirkpatrick}},
  \bibinfo{journal}{J.~Phys.~A.: Math. Gen.} \textbf{\bibinfo{volume}{22}},
  \bibinfo{pages}{L149} (\bibinfo{year}{1989}).

\bibitem[{\citenamefont{Kawasaki and Miyazima}(1997)}]{Kawasaki:97}
\bibinfo{author}{\bibfnamefont{K.}~\bibnamefont{Kawasaki}} \bibnamefont{and}
  \bibinfo{author}{\bibfnamefont{S.}~\bibnamefont{Miyazima}},
  \bibinfo{journal}{Z.~Phys.~B} \textbf{\bibinfo{volume}{103}},
  \bibinfo{pages}{423} (\bibinfo{year}{1997}).

\bibitem[{\citenamefont{Kawasaki}(1994)}]{Kawasaki:94}
\bibinfo{author}{\bibfnamefont{K.}~\bibnamefont{Kawasaki}},
  \bibinfo{journal}{Physica A} \textbf{\bibinfo{volume}{208}},
  \bibinfo{pages}{35} (\bibinfo{year}{1994}).

\bibitem[{\citenamefont{L{\"o}wen}(2003)}]{Loewen:03}
\bibinfo{author}{\bibfnamefont{H.}~\bibnamefont{L{\"o}wen}},
  \bibinfo{journal}{J.~Phys.:~Condens.~Matter} \textbf{\bibinfo{volume}{15}},
  \bibinfo{pages}{V1} (\bibinfo{year}{2003}).

\bibitem[{\citenamefont{Penna and Tarazona}(2006)}]{Penna:06}
\bibinfo{author}{\bibfnamefont{F.}~\bibnamefont{Penna}} \bibnamefont{and}
  \bibinfo{author}{\bibfnamefont{P.}~\bibnamefont{Tarazona}},
  \bibinfo{journal}{J.~Chem.~Phys.} \textbf{\bibinfo{volume}{124}},
  \bibinfo{pages}{164903} (\bibinfo{year}{2006}).

\bibitem[{\citenamefont{Royall et~al.}(2007)\citenamefont{Royall, Dzubiella,
  Schmidt, and van Blaaderen}}]{Royall:07}
\bibinfo{author}{\bibfnamefont{C.~P.} \bibnamefont{Royall}},
  \bibinfo{author}{\bibfnamefont{J.}~\bibnamefont{Dzubiella}},
  \bibinfo{author}{\bibfnamefont{M.}~\bibnamefont{Schmidt}}, \bibnamefont{and}
  \bibinfo{author}{\bibfnamefont{A.}~\bibnamefont{van Blaaderen}},
  \bibinfo{journal}{Phys.~Rev.~Lett.} \textbf{\bibinfo{volume}{98}},
  \bibinfo{pages}{188304} (\bibinfo{year}{2007}).

\bibitem[{\citenamefont{Rex and L{\"o}wen}(2008)}]{Rex:08}
\bibinfo{author}{\bibfnamefont{M.}~\bibnamefont{Rex}} \bibnamefont{and}
  \bibinfo{author}{\bibfnamefont{H.}~\bibnamefont{L{\"o}wen}},
  \bibinfo{journal}{Phys.~Rev.~Lett.} \textbf{\bibinfo{volume}{101}},
  \bibinfo{pages}{148302} (\bibinfo{year}{2008}).

\bibitem[{\citenamefont{van Teeffelen
  et~al.}(2008{\natexlab{b}})\citenamefont{van Teeffelen, L{\"o}wen, and
  Likos}}]{SvenJPCM}
\bibinfo{author}{\bibfnamefont{S.}~\bibnamefont{van Teeffelen}},
  \bibinfo{author}{\bibfnamefont{H.}~\bibnamefont{L{\"o}wen}},
  \bibnamefont{and} \bibinfo{author}{\bibfnamefont{C.~N.} \bibnamefont{Likos}},
  \bibinfo{journal}{J.~Phys.:~Condens.~Matter} \textbf{\bibinfo{volume}{20}},
  \bibinfo{pages}{404217} (\bibinfo{year}{2008}{\natexlab{b}}).

\bibitem[{pfc({\natexlab{a}})}]{pfc_footnote_EMA}
\bibinfo{note}{Note, that the static properties of the dipolar system under
  study, including the equilibrium phase diagram, have been obtained applying
  different and more accurate approximations to the excess free energy
  functional~\cite{Teeffelen:06, SvenJPCM}.}

\bibitem[{\citenamefont{Elder and Grant}(2004)}]{Elder:04}
\bibinfo{author}{\bibfnamefont{K.~R.} \bibnamefont{Elder}} \bibnamefont{and}
  \bibinfo{author}{\bibfnamefont{M.}~\bibnamefont{Grant}},
  \bibinfo{journal}{Phys.~Rev.~E} \textbf{\bibinfo{volume}{70}},
  \bibinfo{pages}{051605} (\bibinfo{year}{2004}).

\bibitem[{\citenamefont{Tupper and Grant}(2008)}]{Tupper:08}
\bibinfo{author}{\bibfnamefont{P.~F.} \bibnamefont{Tupper}} \bibnamefont{and}
  \bibinfo{author}{\bibfnamefont{M.}~\bibnamefont{Grant}},
  \bibinfo{journal}{Europhys.~Lett.} \textbf{\bibinfo{volume}{81}},
  \bibinfo{pages}{40007} (\bibinfo{year}{2008}).

\bibitem[{\citenamefont{Hansen and McDonald}(2006)}]{hansen-mcdonald:86}
\bibinfo{author}{\bibfnamefont{J.-P.} \bibnamefont{Hansen}} \bibnamefont{and}
  \bibinfo{author}{\bibfnamefont{I.~R.} \bibnamefont{McDonald}},
  \emph{\bibinfo{title}{Theory of simple liquids}}
  (\bibinfo{publisher}{Academic Press}, \bibinfo{address}{London},
  \bibinfo{year}{2006}), \bibinfo{edition}{3rd} ed.

\bibitem[{\citenamefont{van Teeffelen et~al.}(2006)\citenamefont{van Teeffelen,
  Likos, Hoffmann, and L{\"o}wen}}]{Teeffelen:06}
\bibinfo{author}{\bibfnamefont{S.}~\bibnamefont{van Teeffelen}},
  \bibinfo{author}{\bibfnamefont{C.~N.} \bibnamefont{Likos}},
  \bibinfo{author}{\bibfnamefont{N.}~\bibnamefont{Hoffmann}}, \bibnamefont{and}
  \bibinfo{author}{\bibfnamefont{H.}~\bibnamefont{L{\"o}wen}},
  \bibinfo{journal}{Europhys.~Lett.} \textbf{\bibinfo{volume}{75}},
  \bibinfo{pages}{583} (\bibinfo{year}{2006}).

\bibitem[{\citenamefont{Rogers and Young}(1984)}]{Rogers:84}
\bibinfo{author}{\bibfnamefont{F.~J.} \bibnamefont{Rogers}} \bibnamefont{and}
  \bibinfo{author}{\bibfnamefont{D.~A.} \bibnamefont{Young}},
  \bibinfo{journal}{Phys.~Rev.~A} \textbf{\bibinfo{volume}{30}},
  \bibinfo{pages}{999} (\bibinfo{year}{1984}).

\bibitem[{pfc({\natexlab{b}})}]{pfc_footnote_s}
\bibinfo{note}{In particular, we chose $s=60$ for $31\leq\Gamma\leq35$, $s=7$
  for $37\leq\Gamma<40$, and $s=5$ for $\Gamma\geq40$, in the DDFT, $s=41$ for
  $\Gamma=33$, $s=21$ for $\Gamma=37$, $s=11$ for $38\leq\Gamma\leq60$, in the
  PFC1 model, and $s=83$ for $\Gamma=37$, $s=41$ for $\Gamma=38$, $s=21$ for
  $\Gamma=39$, and $s=11$ for $38\leq\Gamma\leq60$, in the PFC2 model.}

\bibitem[{\citenamefont{K{\"o}ppl et~al.}(2006)\citenamefont{K{\"o}ppl,
  Henseler, Erbe, Nielaba, and Leiderer}}]{Koppl:06}
\bibinfo{author}{\bibfnamefont{M.}~\bibnamefont{K{\"o}ppl}},
  \bibinfo{author}{\bibfnamefont{P.}~\bibnamefont{Henseler}},
  \bibinfo{author}{\bibfnamefont{A.}~\bibnamefont{Erbe}},
  \bibinfo{author}{\bibfnamefont{P.}~\bibnamefont{Nielaba}}, \bibnamefont{and}
  \bibinfo{author}{\bibfnamefont{P.}~\bibnamefont{Leiderer}},
  \bibinfo{journal}{Phys.~Rev.~Lett.} \textbf{\bibinfo{volume}{97}},
  \bibinfo{pages}{208302} (\bibinfo{year}{2006}).

\bibitem[{pfc({\natexlab{c}})}]{pfc_footnote_steadystate}
\bibinfo{note}{The discussion of the behavior of $v_f(t\gg\tau_B)$ in the DDFT
  and the PFC models is deferred to a future publication. For the same behavior
  in a similar model of hard spheres, see~\cite{Wild:01}.}

\bibitem[{\citenamefont{Wild and Harrowell}(2001)}]{Wild:01}
\bibinfo{author}{\bibfnamefont{R.}~\bibnamefont{Wild}} \bibnamefont{and}
  \bibinfo{author}{\bibfnamefont{P.}~\bibnamefont{Harrowell}},
  \bibinfo{journal}{J.~Chem.~Phys.} \textbf{\bibinfo{volume}{114}},
  \bibinfo{pages}{9059} (\bibinfo{year}{2001}).

\end{thebibliography}
\end{document}